\documentstyle[prl,aps]{revtex}

\newcommand{\BE}{\begin{equation}}
\newcommand{\EE}{\end{equation}}
\newcommand{\BA}{\begin{eqnarray}}
\newcommand{\EA}{\end{eqnarray}}
\begin{document}
\draft

\twocolumn[\hsize\textwidth\columnwidth\hsize\csname@twocolumnfalse\endcsname

\title {Coherence in a coupled network:
Implication for brain function}
\author{Zhen Ye}
\address{Wave Phenomena Laboratory, Department of Physics, National Central University, Chungli, Taiwan 32054}

\date{Final version of physics/0005076} \maketitle

\begin{abstract}

In many body systems, constituents interact with each other,
forming a recursive pattern of mutual interaction and giving rise
to many interesting phenomena. Based upon concepts of the modern
many body theory, a model for a generic many body system is
developed. A novel approach is used to investigate the general
features in such a system. An interesting phase transition in the
system is found. Possible link to brain dynamics is discussed. It
is shown how some of the basic brain processes, such as learning
and memory, find therein a natural explanation.

\end{abstract}

\pacs{PACS numbers: 05.70.Fh, 05.45.Xt, 87.10.+e} ]

A major task in physics research is to search for general
principles behind various nature processes. One of the greatest
advances in the last century is understanding of how a many body
system manifests and behaves, leading to many important phenomena
including the multiple scattering of both quantum and classical
waves, the quantum Hall effect, the transition between
super-conducting and normal metal states and so on. It has become
a daily experience that macroscopic objects arise from interaction
between individuals in a many body system which may be of
microscopic scale.

In this way, we were led to a fundamental question: are there
general features which can be said about the manifestation of many
body interactions. Addressing this question, many profound
concepts have been developed and matured over the past half
century. To name a few, this includes the phase transition,
invariance or symmetry, and symmetry breaking; these concepts
fabricate the quilt of the modern many body theory\cite{Umezawa}.
An intricate inter-connection among these is stated by the
Goldstone theorem. A symmetry breaking will result in not only a
phase transition, but also the appearance of some sort of long
range order, or synchronization, or coherence, depending on what
subject is under scrutiny. For example, the phase transition
between the liquid and solid states is a result of breaking the
continuous space invariance. Once the symmetry breaks, phonon
modes appear to reflect the collective behavior of the system.
Although they can be systematically obtained in the realm of the
quantum field theory, for instance, from the Ward-Takahashi
relation, these phonon modes are not of quantum origin. This
observation leads us to believe that the long-range macroscopic
coherence or order does not necessarily require a quantum cause.

Being a most complicated classical hierarchy kingdom, human brain
functions in such a way that it apparently involves many
interacting elements. Natural questions may arise as whether the
concepts of the many body theory can be applied to the problem of
brain, and what general features can be depicted about the brain
function. Physical research on brain function is not merely out of
the scientific curiosity, but rather the problem is one of the
most basic research subjects and has long been an outstanding
inquiry from various scientific communities.

In view of the complexity of a brain, any task on brain function
seems unfeasible at first sight. Towards this, we may draw
attention to the study of gas systems. Even in a simplest perfect
gas system, it is absolutely unrealistic to trace the trajectory
of any individual gas molecules in the realm of Newton's
mechanics. Such a seemingly intractable problem, however, has been
studied successfully in terms of statistical mechanics. Therefore,
whether a most difficult problem can be formulated and has certain
answers crucially depends on from which angle and at what level
the problem is investigated. Searching for a general description of nature
phenomena without having to tangle in the mist of complexity and to
know the details of each consisting component in complicated systems is in fact
one of the major tasks for physicists.

Motivated by our recent work on wave localization\cite{Ye}, here
we formulate a general model for interacting many body systems. It
will be clear that this model can represent many important
physical situations, and it may also be applicable to model brain
systems. The moral is that as long as there is certain consensus
about the brain function, a general underlying principle can be
identified. Then a new approach will be proposed to study the
model. Numerical examples will be presented, then followed by a
discussion on their implications for brain function. It will
become clear that some of the basic brain processes, such as
learning and memory, find therein a natural explanation.
Undeniably, the present investigation has to be primitive due to
limited information and the type of information available about
the brain. Nevertheless, the study does provide a most general
insight to the problem of brain function with respect to features
involving many body interaction.

Without compromising generality and for the sake of simplicity, we
start with considering a system consisting of $N$ electric
dipoles, in the hope of setting up a generic model for many body
systems. These dipoles can be either {\it randomly} or {\it
regularly} distributed in a space which can be either one, two, or
three dimensional. The former forms random arrays of objects,
while the latter would form a crystal lattice allowing band
structures to appear. The dipoles interact with each other through
transmitting electrical waves when each dipole oscillates.

The equation of motion for any dipole can be easily deduced from
the electro-magnetic theory\cite{textbook}. Neglecting terms
higher than the second order, for the $i$-th dipole, for example,
the equation of motion for the dipole oscillation is \BA
&&\frac{d^{2}Q_{i}}{dt^2}+
\omega^{2}_{i}Q_{i}+\gamma_i\frac{dQ_{i}}{dt}= F_i(t)\nonumber\\
&+& \sum_{j=1, j\neq i}^{N} C_{ij}\left(1 -
\frac{1}{2}Q_i\hat{z_i}\cdot\nabla\right)
\frac{{D^{ij}}}{Dt^2}\frac{{Q}_j(t - |\vec{r}_i -
\vec{r}_j|/c)}{|\vec{r}_i-\vec{r}_j|}, \nonumber\\ \label{eq:1}
\EA where $$ \frac{D^{ij}}{Dt^2} =
[(\hat{r}_{ij}\cdot\hat{z}_i)(\hat{r}_{ij}\cdot\hat{z}_j)-1]\frac{d^2}{dt^2}.$$
In these, $Q_i$ is the oscillation displacement with $\omega_i$
being the natural frequency and $\gamma_i$ being the damping
factor, $\hat{z}_i$ is the unit vector along the dipole direction
of the $i$-th dipole, $\hat{r}_{ij}$ is the unit vector pointing
from the $j$-th dipole to the $i$-th dipole, the coupling constant
$C_{ij} = \frac{-q_iq_j\mu_0}{2\pi m_i}$. In the present paper we
ignore the rotational motion of the dipoles, which can be added
with ease.

The second term on the RHS of Eq.~(\ref{eq:1}) denotes the
nonlinear interaction. For the first step analysis, this effect
may be dropped. Equation (\ref{eq:1}) can formally be generalized
as \BE {\cal L}_i(\partial) \psi_i = F_i + \sum_{j=1,j\neq i}
\hat{T}_{ij} [\psi_j], \label{eq:2} \EE where $\psi_i$ is the
physical quantity under consideration; in the dipole case $\psi_i$
is the displacement, $F_i$ refers to the external stimuli at the
$i$-th entity, $\hat{T}_{ij}$ is an operator transferring
interaction between the $i$-th and $j$ component which usually can
be written in terms of a Bosonic propagator.  The operator ${\cal
L(\partial)}$ is called divisor obtained when the many body
interaction is not present\cite{Umezawa}. So far, in nature there
are two known types of divisors: \BA \mbox{Type I}: \ {\cal
L(\partial)} &=& \frac{1}{c^2}\frac{\partial^2}{\partial t^2} -
\omega^2(\nabla),\nonumber\\ \mbox{Type II}: \ {\cal L(\partial)}
&=& i\hbar \frac{\partial}{\partial t} - \epsilon(\nabla).
\nonumber\EA Type I usually refers to the boson like systems,
while Type II divisor usually refers to the fermion like systems.
In the above $\omega(\nabla)$ and $\epsilon(\nabla)$ refer to the
operators that give the energy spectra in absence of interactions,
i.~e. the bare energy.  When dissipation is present, a term like
$\gamma\frac{\partial}{\partial t}$ should be added to the
divisors.

Equation (\ref{eq:2}) accommodates a variety of problems of great
interest. To name but a few, the vibration of the lattice formed
by atoms is described and multiple scattering of waves in a system
consisting of many scatterers are described by a version of Type I
equation. For an electronic system, in terms of the Local
Combination of Atomic Orbits, the behavior of the system is
governed by a Type II equation. Other examples can be referred to
in \cite{disorder}.

In short, Eq.~(\ref{eq:2}) represents a wide range of many body
problems. It reflects a basic many body interaction picture. The
interaction among the constituents is mediated by waves. This is
common in nature. We would like to explore some new features
embedded in Eq.~(\ref{eq:2}). We consider wave propagation in a
Type I system. The incident wave will be interfered by radiated
waves from each entity. The energy flow is $\vec{J}\sim
\mbox{Re}[u^\star(-i\nabla)u].$ Writing the field as
$Ae^{i\theta}$, the current becomes $A^2\nabla\theta$, a version
of the Meissner equation. This hints at that when $\theta$ is
constant while $A\neq 0$, the flow stops and energy will be
localized or stored in space. This phase transition is
condensation of modes in the real space, in contrast to the Cooper
pair condensation in the momentum space in the superconductivity.
This is useful as it implies that energy can be stored in certain
spatial domians; as will be clear, it links to some basic features
of brain function.

We solve Eq.~(\ref{eq:2}) for Type I situations in the frequency
domain. There are $N$ bodies in the system,
and they are excited by a point source located
in the middle of these bodies. To further simplify our discussion,
we assume that all elements in the system are identical and the
interaction is isotropic; the study of anisotropic cases is
possible but comuptationally expensive and some examples have been
documented elsewhere\cite{AA}. Explicitly, from Eq.~(\ref{eq:1})
the governing equation can be written as \BA
\frac{d^2}{dt^2}\psi_i &+& \omega_0^2\psi_i + \gamma\frac{d}{d
t}\psi_i = F_i\nonumber\\ &+& \sum_{j=1,j\neq i}^N \left.C
\frac{\ddot{\psi_i}}{|\vec{r}_i-\vec{r}_j|^{(d-1)/2}}\right|_{t-|\vec{r}_i-\vec{r}_j|/c},
\EA where $C$ is the coupling constant, and $c$ is the wave speed.
Strictly, for 2D ($d=2$), the propagator should be in a form of
the zero-th order Hankel function of the first kind\cite{Ye}.

We write \BE \psi_i = A_ie^{i\theta_i},\EE where we dropped the
common time factor $e^{-i\omega t}$. The total wave in the space
is the summation of the direct wave from the source and the
radiated waves from all the bodies. For each phase $\theta_i$, we
define a phase vector such that \BE \vec{v}_i=\cos\theta_i
\vec{e}_x + \sin\theta_i\vec{e}_y.\EE This phase vector can be
placed on the respective entity and put on a 2D plane spanned by
$\vec{e}_x, \vec{e}_y$. The amplitude $A_i$ refers to the energy
distributed among the entities.

Numerical computation has been carried out for 1D, 2D and 3D
situations respectively. The results are summarized as follows.
(1) In the 2 and 3 dimensional cases, for sufficient large number
density of bodies, e.~g. exceeding roughly $1/C^2$ in 3D, and for
small damping factor, the energy is trapped and stored inside the
system for a range of frequencies located above the nature
frequency. More explicitly, at a frequency in that range, the
transmitted wave from the exciting source is localized inside the
system. There is no energy flow, as supported by the appearance of
a global coherence in the phase vectors defined above, i.~e. all
phase vectors point to the same direction and phase $\theta$ is
constant. (2) When the damping increases, transferring more and
more energy to other forms, the coherence phenomenon starts to
disappear gradually. (3) Decreasing the density of the
constituents also allows the energy to be released from the
system, meanwhile the coherence disappear. Another effect of
decreasing density is to narrow the trapping frequency range. (4)
Outside the trapping frequency range, no energy can be stored in
the system. Therefore adjusting the frequency will crucially tune
whether the system can store energy. (5) For 1D cases, energy is
stored near the source for all frequencies with any given amount
of random placement of the entities. (6) Last, perhaps most
importantly, all these features are not sensitive to any
particular distribution of the constituents. In other words, these
features remain unchanged as the configuration of the system is
varied.

These features are illustrated by a 2D situation shown in Fig.~1
Here is shown that no energy is stored for $f/f_0=0.98039$ and
5.8824, but the energy becomes trapped for $f/f_0=1.9608$. Clearly
there are phase transitions from $f/f_0=0.98039$ to 1.9608, and
from 1.9608 to 5.8824. We stress that these features do not
require quantum origin. Here $f_0$ refers to the nature frequency.
It is also clear from Fig.~1 that when the energy is stored, a
striking coherence behavior appears in the system. Similar
features are also observed for one and three-dimensional
situations.

In order to characterize the phase transition, we define an order
parameter as $\Delta = \frac{1}{N}\left|\sum_{i=1}^N
\vec{v}_i\right|$. The magnitude of the order parameter represents
the degree of coherence in the system. Obviously, when all the
phase vectors are pointing to the same direction, the order
parameter will equal one. The order parameter is plotted for one
random realization as a function of the reduced frequency $f/f_0$
in Fig.~2. The two phase transitions appear around $f/f_0 = 1.6$
and $3.6$. The apparent fluctuations below $f/f_0 = 1.5$ and above
$4$ are due to effects of random distribution and finite sample
size.

Now we are ready to discuss the linkage to brain function.
Although brain may be the most complex system in the nature, still
several general observations (dogmas) can be stated. First, brain
is a system of many body. Any single neuron should not be
significant for the whole brain, but rather the patterns of
activity of clusters are important. As early as in 1967, Ricclardi
and Umezawa first proposed to explain the brain dynamics in the
context of many body theory\cite{LM}. Anderson\cite{Anderson} also
discussed some concepts in the many body theory with connection to
living matter. Their work has directly or indirectly stimulated
and continues to inspire much research along the line (e.~g.
\cite{Hameroff,Arani,Vitiello,Sivakami}). A special area in brain
research has emerged and is known as quantum brain
dynamics\cite{QBD}. Second, experiments have indicated that there
are coherence phenomena in brain function. Neurons or neuron
clusters at different spatial points act collectively, performing
a specific task. Such coherence is also observed for microtubles
in nerve cells\cite{Jibu} and other biological systems\cite{F}. To
effectively performing long range collective tasks requires the
communication to be mediated by waves and the brain system to be
boson-like. Third, brain not only accumulates but stores energy,
corresponding to learning and memory processes respectively. As
memories in a healthy brain are flashed in a regular pattern, it
is reasonable to believe that the memory as a form of energy is
stored in a coherent manner. Fr\"olich was the first to link
energy storage with long range coherence in certain biological
structures\cite{F}. Fourth, the brain function can resist
reasonable damages which may cause the brain structure to
dislocate to a rational extent. This implies that the brain
function can be executed in relatively disordered environments.
Lastly, brain can age, losing memory and becoming dying.

All these important features of brain can be well explained in the
realm of the above generic model. Although it is not possible to
discern the exact entities that carry out brain function at this
stage, a few possibilities have been mentioned in the literature,
such as neurons or corticons, proteins in microtubules and so on.
Even water molecules are also mentioned as a core role in
memorizing and storing\cite{PRL}. Putting details aside, what is
important is only a {\it quantity} somehow related to the activity
of brain. Actually, it is known that neurons do not seem to be the
only fundamental units of the brain; or, at least, together with
them some other elements, as for instance {\it glia celles}, may
play an important role\cite{SM}. No matter what they exactly are,
the necessity for stability requires the constituents to follow
Type I equation. The communication between entities should be
mediated by waves to ensure effective information exchange. Based
on the above results, we argue that the brain consists of many
subsets of Type I entities, each subset has a different natural
frequency, and accomplishes different duties. Different subsets
will store specific information depending on the spectra of
external stimuli. Once the stimuli fits into a trapping range of a
specific subset, the energy of characters of the stimuli will be
stored or imprinted into that subset in a coherent manner, as
indicated by Fig.~1. In other words, different subsets response to
respective stimuli and store energy accordingly, corresponding to
learning and memory processes. When the density of entities is not
high enough or there is no enough active entities, however, the
energy cannot be stored for long time. This reveals as the
situation of short memory; this may be seen, for example, in the
early development of children's brains. Long term memory requires
a sufficient amount of active entities and reasonable supply of
stimuli. Moreover, as mentioned above, the feature of long range
coherence and energy storage remains unchanged as the entity
configuration varies; here we exclude the extreme case that all
entities collapse into each other. This would explain the brain's
resistance to reasonable damages. Damages may reduce the active
entities and alter the structure of the system. Either varying the
number of active entities or the natural frequency of entities or
both produces the different states of the brain and controls the
energy release. Memories may also be lost and disoriented when the
the damping effect becomes prominent. The possible contributions
to the damping effect include those from aging effect and
diseases.

In summary, while there are many unanswered questions pertinent to
brain function, necessitated by limited knowledge about brain, it
is almost certain that the brain is a many body system which can
accumulate and store energy. When needed, the stored energy can be
released to accomplish missions. The communication between
different parts in brain is likely mediated by certain brain
waves, as no other means can convey information more effectively
than waves. As long as these general discernments hold, given its
generality the present model expects to play a role in guiding or
stimulating further experimental and theoretical studies.

\vspace{12pt} \noindent The work received support from National
Science Council through the grants NSC-89-2112-M008-008 and
NSC-89-2611-M008-002. Emile Hoskinson is thanked for programming
and useful discussion.

\section*{Figure Caption}

\begin{description}
\item[Figure 1] Energy distribution and diagrams for the phase vectors in a
2D random configuration. Right: Energy distribution; the
geometrical factor has been scaled out. Left: the phase diagrams
for the phase vectors.

\item[Figure 2] Order parameter versus $f/f_0$.

\end{description}

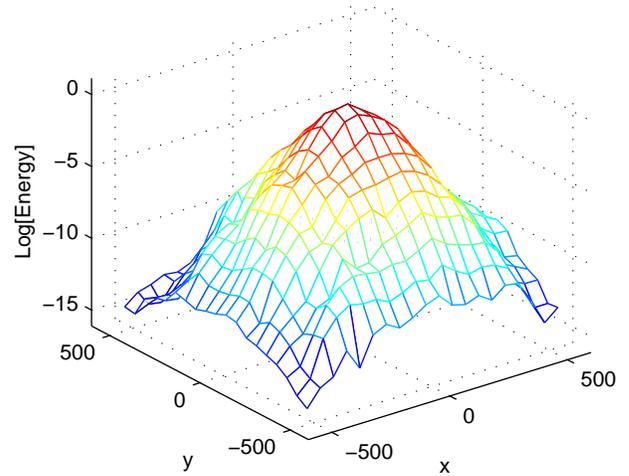
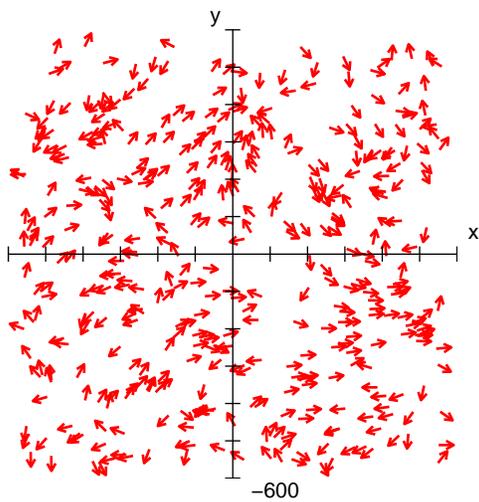
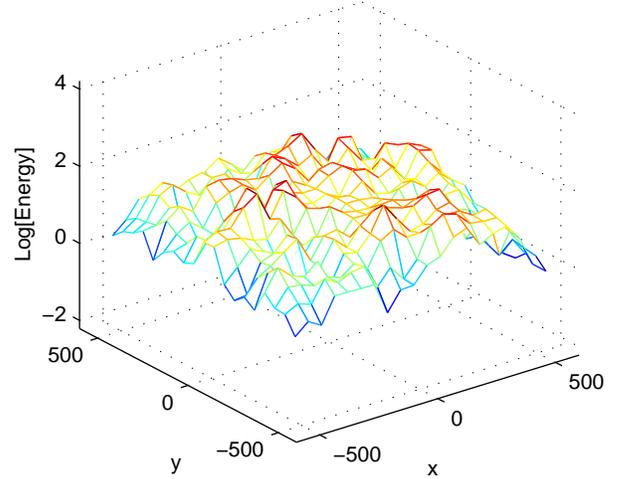
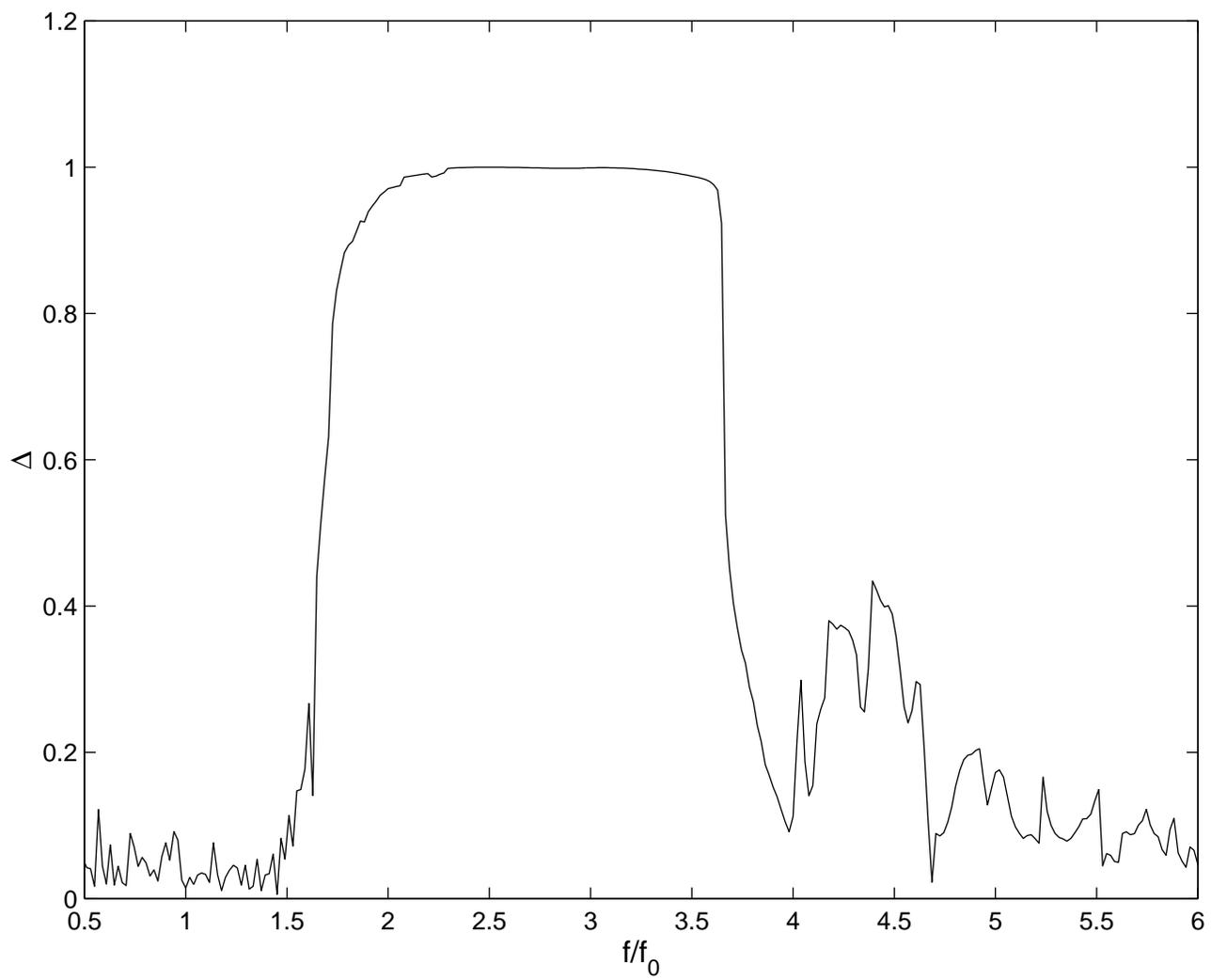
\end{document}